\documentclass[sigconf]{acmart}

\usepackage{graphicx}
\usepackage{amsmath}
\usepackage{multirow}
\usepackage{hyperref}
\usepackage{array}
\hypersetup{colorlinks=true, linkcolor=blue, citecolor=blue, urlcolor=blue}

\AtBeginDocument{%
  \providecommand\BibTeX{{%
    \normalfont B\kern-0.5em{\scshape i\kern-0.25em b}\kern-0.8em\TeX}}}

\acmConference[]{}{}{}
\renewcommand\footnotetextcopyrightpermission[1]{}

\title{Convolutions Need Registers Too: HVS-Inspired Dynamic Attention for Video Quality Assessment}

\author{Mayesha Maliha R. Mithila}
\affiliation{%
  \institution{Texas State University}
  \city{San Marcos}
  \state{Texas}
  \country{United States}
}
\email{mayesha@txstate.edu}

\author{Mylene C.Q. Farias}
\affiliation{%
  \institution{Texas State University}
  \city{San Marcos}
  \state{Texas}
  \country{United States}
}
\email{mylene@txstate.edu}

\begin{abstract}
No-reference video quality assessment (NR-VQA) estimates perceptual quality without a reference video, which is often challenging. While recent techniques leverage saliency or transformer attention, they merely address global context of the video signal by using static maps as auxiliary inputs rather than embedding context fundamentally within feature extraction of the video sequence. We present Dynamic Attention with Global Registers for Video Quality Assessment (\textit{DAGR-VQA}), the first framework integrating \textit{register-token} directly into a convolutional backbone for spatio-temporal, dynamic saliency prediction. By embedding learnable register tokens as global context carriers, our model enables dynamic, HVS-inspired attention, producing temporally adaptive saliency maps that track salient regions over time without explicit motion estimation. Our model integrates dynamic saliency maps with RGB inputs, capturing spatial data and analyzing it through a temporal transformer to deliver a perceptually consistent video quality assessment. Comprehensive tests conducted on the LSVQ, KonVid-1k, LIVE-VQC, and YouTube-UGC datasets show that the performance is highly competitive, surpassing the majority of top baselines. Research on ablation studies demonstrates that the integration of register tokens promotes the development of stable and temporally consistent attention mechanisms. Achieving an efficiency of 387.7 FPS at 1080p, DAGR-VQA demonstrates computational performance suitable for real-time applications like multimedia streaming systems.

\end{abstract}



\begin{document}

\maketitle

\section{Introduction}

With the proliferation of user-generated video content, the ability to efficiently and accurately assess visual quality has become critical for applications such as streaming platforms, video compression systems, and content delivery networks.  Traditional full-reference (FR) video quality assessment (VQA) techniques rely on access to undistorted reference videos, which is often impractical in real-world deployments. This limitation has spurred an increasing interest in no-reference (NR) VQA methods, which estimate perceptual quality directly from processed (possibly distorted) video sequences~\cite{wu2021long, huo2024multi, ruan2021no}. Despite progress, NR-VQA continues to pose significant challenges, as it must effectively model complex and heterogeneous spatio-temporal distortions while closely aligning with subjective quality scores~\cite{yuan2021deep}. Moreover, modern streaming infrastructures demand scalable quality assessment solutions that can operate within computational constraints while maintaining accuracy in various types of content and delivery conditions~\cite{wu2022fastvqa}.

The human visual system (HVS) is an intricate biological system composed of ocular and neurological components that enables the processing of visual stimuli. This system plays a crucial role in guiding cognitive attention towards perceptually \textit{salient} regions while simultaneously filtering out extraneous information. The field of visual saliency prediction draws inspiration from HVS, employing computational models to identify salient regions that inherently attract human attention within images or videos ~\cite{wang2007video, borji2012state}. Salient regions are typically characterized by significant motion, pronounced color contrast, or the presence of distinctive objects. These models have become valuable for VQA, with earlier works assessing quality by combining hand-crafted spatio-temporal features and visual saliency maps to weigh noticeable distortions~\cite{akamine2014video, farias2012performance, vu2011stmad}. However, this approach lacks the adaptability and temporal modeling capabilities of recent deep learning (DL) models.  

SGDNet~\cite{yang2019sgdnet} employs a DL framework that integrates saliency-guided weighting using precomputed saliency maps for image quality assessment. The focus on static saliency within this architecture renders it inadequate for video applications, where visual attention dynamically changes over time. More recent models, like HVS-5M~\cite{zhang2023spatial}, incorporate visual saliency at the initial stages of feature extraction by employing modules focused on content and edge dependency. Nevertheless, the HVS-5M static modular approach lacks dynamic temporal coherence across video sequences.  In contrast, dynamic saliency approaches explicitly model how human focus evolves over time—tracking moving objects, actions, and transitions that capture the viewer's gaze \cite{mital2011clustering, jiang2018deepvs}. The incorporation of temporal awareness is crucial in the evaluation of video quality, as motion and evolving events constantly redirect attention. Most current methods superficially address dynamic attention in later fusion stages, instead of integrating it into the core of feature extraction.

\begin{figure*}[hbt]
    \centering
    \includegraphics[width=\textwidth]{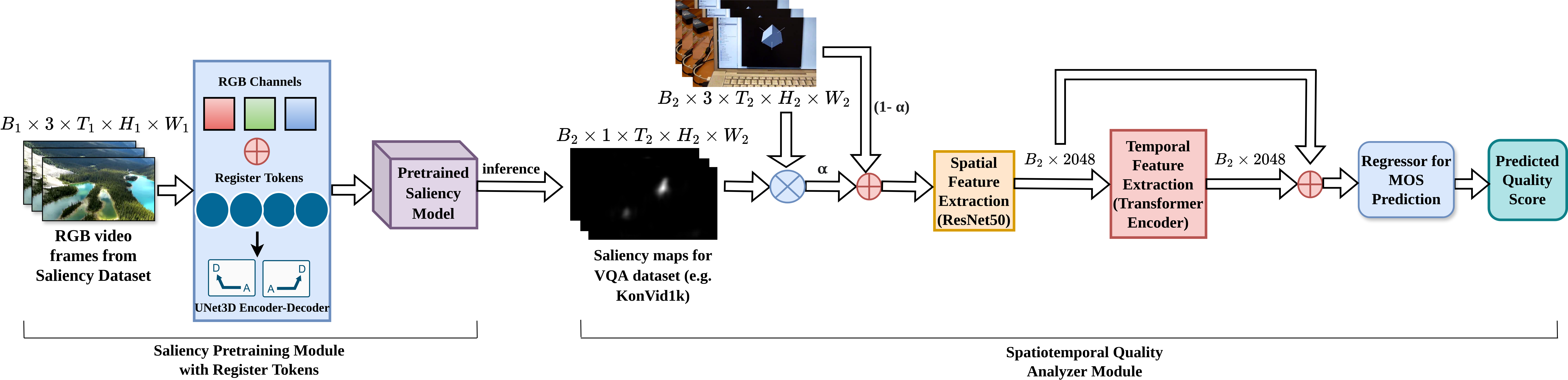}
    \caption{
{Overview of DAGR-VQA. During pretraining, RGB frames from saliency dataset ($B_1 \times 3 \times T_1 \times H_1 \times W_1$) are concatenated ($\bigoplus$) with register-tokens and processed by a UNet3D. The pretrained model inferences saliency maps ($B_2 \times 1 \times T_2 \times H_2 \times W_2$) for VQA videos. For each frame, fusion is performed as $(1-\alpha)$ times original frames, added ($\bigoplus$) with $\alpha$ times saliency map element-wise multiplied ($\bigotimes$) original frames. Spatial, temporal, and regression modules complete the VQA pipeline.}
}
    \label{fig:DAGR-VQA}
\end{figure*}
Recent advancements in transformer-based architectures, along with the implementation of register-tokens, have demonstrated significant potential in effectively capturing global context within video signals~\cite{dosovitskiy2020image, darcet2024registers}. Register-tokens address a key limitation of standard attention by offering dedicated memory for global information. By contrast, classification (CLS) tokens must support both classification and representation, while patch tokens remain inherently local. Explicit modeling of scene-level memory is critical for VQA because a consistent global context allows the model to measure perceived quality over time. However, in current NR-VQA pipelines, register-tokens are usually restricted to transformer branches and are rarely used in convolutional backbones. Earlier NR-VQA models  struggled to capture how human attention shifts with content, often relying on expensive motion analysis (e.g., optical flow)~\cite{borji2012state, culibrk2011salient}. Later models, like KVQ, integrate region-specific dynamic saliency into transformer attention~\cite{qu2025kvq}.  FineVQ adds multimodal fusion of video and text~\cite{duan2025finevq}. These approaches lack HVS-inspired dynamic saliency in feature extraction, prompting the integration of global context priors like register-tokens into the backbone to connect dynamic attention with efficient spatio-temporal representations for high-performance video quality assessment.

In this work, we present Dynamic Attention with Global Registers for Video Quality Assessment (DAGR-VQA). The overall framework of DAGR-VQA is shown in Figure~\ref{fig:DAGR-VQA} and consists of a register-token-augmented convolutional backbone for extracting salient dynamic features, a spatial encoder, and a lightweight temporal transformer.  We embed register-tokens~\cite{darcet2024registers} as explicit global priors within a convolutional UNet3D backbone, extending context aggregation to the video domain. Register-tokens facilitate the integration of scene-level information across multiple frames while preserving temporal coherence. This process emulates the HVS) by directing attention to salient regions as scenes progress. Unlike previous NR-VQA methods that rely on transformer modules~\cite{xing2022starvqa, Wu2023Dover} or explicit motion cues~\cite{jiang2018deepvs, duan2025finevq}, our architecture fuses adaptive saliency with lightweight temporal modeling for accurate real-time predictions. The main contributions are as follows.
\begin{itemize}
\item We present a pretraining saliency framework with register-tokens embedding global priors during convolutional feature extraction. This method mimics dynamic, HVS-inspired attention without self-attention modules or motion estimation like optical flow.
\item We design a modular end-to-end NR-VQA framework where register-tokens unify global context and quality prediction directly at the backbone level, unlike prior approaches that use saliency and context only peripherally.
\item We validate our methodology on four key NR-VQA benchmarks, demonstrating competitive performance. We also conduct ablation studies on register-tokens and architectural components, highlighting their role in developing stable, temporally coherent attention mechanisms.
\end{itemize}

\section{Related Work}

Advances in deep learning have improved spatio-temporal modeling for VQA, enhancing prediction accuracy. However, strong generalization remains challenging due to dataset biases and limitations in temporal modeling~\cite{yuan2021deep,zhang2020no}. Recent methods explore spatio-temporal attention~\cite{chen2022spatiotemporal} and perceptual or saliency-based cues~\cite{zhao2022perceptual}, yet these can be computationally expensive and lack robustness. In response, deep feature fusion, advanced attention modules, and transfer learning techniques have been studied. For example, ReLaX-VQA~\cite{wang2024relaxvqa} uses residual fragments and layered feature stacking to strengthen spatio-temporal pattern perception, although without explicit saliency fusion. Li \textit{et al.}~\cite{li2022unique} demonstrated that quality-aware pre-training enhances robustness to video distortions, highlighting the value of eye-tracking and perceptual datasets.

Incorporating HVS principles into VQA ensures that quality predictions align with human perception rather than purely algorithmic metrics. Although early approaches relied on hand-crafted features, deep neural networks have substantially improved spatio-temporal saliency prediction. For example, DeepVS~\cite{jiang2018deepvs} combines object and motion cues, while ViNet-S~\cite{girmaji2025vinet} uses a lightweight UNet-based architecture guided by action information. However, such models still often rely on explicit motion cues or static features, limiting their ability to adapt to diverse, dynamic shifts in human attention, a limitation particularly important given that eye-tracking studies show editing and cinematography shape viewer attention~\cite{bruckert2023movies}, motivating approaches that integrate dynamic saliency prediction as a core, learnable mechanism.

Predicted saliency has become increasingly used as a cue for VQA performance assessment~\cite{akamine2014video, zhang2017study}. Early approaches like ST-MAD~\cite{vu2011stmad} weighed distortions by perceptual salience using predefined, rule-based maps. Most early methods rely on static saliency, computing attention maps per frame without temporal modeling, whereas dynamic saliency tracks attention shifts across scenes—crucial for video, where attention is inherently time-dependent. Later models like SGDNet~\cite{yang2019sgdnet} and HVS-5M~\cite{zhang2023spatial} incorporate saliency but still lack full temporal coherence. More recently, KVQ~\cite{qu2025kvq} uses Fusion-Window Attention for saliency extraction and advances temporal modeling. However, these approaches treat saliency as auxiliary or late-stage rather than embedding it fundamentally in the backbone. This reflects both architectural choices and the scarcity of large-scale datasets with frame-level saliency ground truth~\cite{alers2012examining, bruckert2023movies}, hindering backbone-integrated dynamic saliency development.

Effective spatio-temporal modeling is essential for NR-VQA, as static feature extraction alone cannot account for perceptual degradations that unfold over time. Freitas \textit{et al.}~\cite{freitas2018using} propose a video quality metric that integrates texture, saliency, spatial activity, and temporal features through a random forest regression model. 
Vu \textit{et al.}~\cite{vu2011stmad} presented ST-MAD, a model that captures spatio-temporal distortion by emphasizing perceptually important regions in motion sequences. This model relies on hand-crafted features and lacks the fine-grained spatio-temporal alignment that more recent deep learning approaches strive to achieve. Building on this, Ying \textit{et al.}~\cite{ying2021patchvq} proposed PatchVQ, which uses a bidirectional LSTM to learn rich spatio-temporal representations from video patches. More recently, Hu \textit{et al.}~\cite{hu2024video} investigated the efficiency–effectiveness trade-off in VQA by introducing joint spatial and temporal sampling, demonstrating that heavily downsampled videos can still yield competitive predictions with a lightweight online VQA. Collectively, these works underscore the necessity of spatio-temporal modeling. 

 Wu \textit{et al.}~\cite{wu2022fastvqa} introduced FAST-VQA with Grid Mini-patch Sampling (GMS), an adaptive mechanism that improves computational efficiency while preserving global quality assessment through uniformly distributed sampling. Morra \textit{et al.}~\cite{morra2021variability} explored how dataset biases impact quality assessment predictions in user-generated content. Sun \textit{et al.}~\cite{sun2024analysis} demonstrated that the trade-off between model complexity and simplicity depends on dataset composition, and that adaptive feature selection methods enhance spatio-temporal extraction while maintaining computational efficiency, a critical necessity for resource-constrained VQA systems.

Visual transformers, originally designed for images~\cite{dosovitskiy2020image}, have been adapted for video tasks~\cite{arnab2021vivit} and are now fundamental for modeling temporal dependencies in VQA. MS-SCANet~\cite{mithila2025ms-scanet} is a multiscale NR-IQA model that combines spatial and channel features using cross-branch attention, but lacks frame-level temporal modeling for images. Video-centric architectures, such as ViViT~\cite{arnab2021vivit}, implement temporal attention mechanisms to capture motion dynamics, whereas STARVQA~\cite{xing2022starvqa} employs an integrated spatial–temporal attention approach. Self-attention performs well but has a quadratic cost with sequence length, memory limits in complex tasks, and sensitivity to anomalies. DOVER~\cite{Wu2023Dover}, a top transformer-based NR-VQA model, features local and global attention for cross-dataset generalization, though its global context remains limited to attention layers. FineVQ~\cite{duan2025finevq} improves NR-VQA with multimodal fusion and LLMs for attribute-aware evaluations, offering interpretable quality scores but downplays dynamic saliency and global priors at the backbone level. Darcet \textit{et al.}~\cite{darcet2024registers} introduced register-tokens as trainable memory units to preserve global information. Unlike CLS tokens that are as classification heads and local patch tokens, register-tokens specialize in scene-level context and have been used primarily in large transformer systems and generic vision tasks, rather than lightweight CNN backbones or perceptual VQA. 

\begin{figure*}[tb]
    \centering
    \includegraphics[width=.95\textwidth]{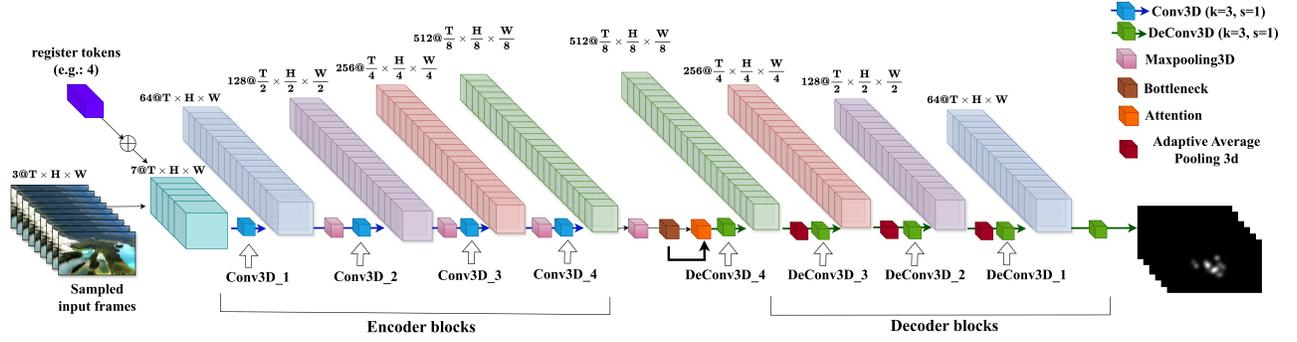}  
    \caption{
    Saliency pretraining architecture with register-tokens. Sampled input frames are concatenated ($\oplus$) with learnable register-tokens and processed by a UNet3D-style encoder-decoder. The network applies successive 3D convolutions, pooling, a bottleneck, attention, and adaptive pooling to produce temporally stable, quality-aware saliency maps. Channel and spatial dimensions at each stage are indicated above the blocks.}
    \label{fig:saliency_pretraining}
\end{figure*}
\section{Methodology}

Our architecture, shown in Figure~\ref{fig:DAGR-VQA}, has two main modules: Saliency Pre-training with Register-Tokens and a Spatio-temporal Video Quality Analyzer. These components are described next.  

\subsection{Saliency Pretraining with Register-Tokens}

We employ a UNet3D encoder-decoder architecture enhanced with register-tokens—learnable global priors motivated by Vision Transformers~\cite{darcet2024registers} (Figure \ref{fig:saliency_pretraining}). These tokens are incorporated directly within the convolutional backbone, allowing the model to capture global context without the need for self-attention or recurrent mechanisms. Unlike approaches limited to static image saliency, such as SGDNet~\cite{yang2019sgdnet}, our framework explicitly learns to predict spatio-temporal saliency. In contrast to ViNet-S~\cite{girmaji2025vinet}, our architecture achieves a more precise modeling of motion dynamics and spatial detail through the extensive use of 3D convolutions along with register-tokens. We train the saliency network entirely from scratch on the DHF1K dataset~\cite{wang2018revisiting}, ensuring that register-tokens are seamlessly integrated into the full spatio-temporal pipeline. Existing pre-trained solutions like TASED-Net~\cite{min2019tased} are often limited by their architectural constraints, making them less suitable for register-token-based models. Although DHF1K is of moderate size, its high-quality eye-tracking annotations provide a solid foundation for supervised training.

\paragraph{Initialization and Concatenation}
Register-tokens act as learnable global priors that are concatenated with the input to bias the encoder toward consistent spatio-temporal patterns. Let $N$ be the number of register-tokens and $d$ their embedding dimension. The register token tensor $\mathbf{R}$ is initialized as:
\begin{equation}
    \mathbf{R} \sim \mathcal{N}(0, 1), \quad \mathbf{R} \in \mathbb{R}^{1 \times N \times d \times 1 \times 1},
\end{equation}
where $\mathcal{N}(0, I)$ denotes the standard normal distribution and the tensor is initialized randomly. The registration tokens are projected to $\mathbf{R}'$ to match the spatio-temporal dimensions of the input using the following 3D convolution:
\begin{equation}
    \mathbf{R}' = \mathcal{T}(\mathbf{R}) = \text{Conv3D}(\mathbf{R}), \quad \mathbf{R}' \in \mathbb{R}^{N \times T \times H \times W},
\end{equation}
where $T$ is the number of frames, $H$ and $W$ are height and width, and $\mathcal{T}$ denotes the 3D convolutional transformation.

The projected register-tokens are concatenated with the video input tensor as follows. The augmented input $\mathbf{V}_{\text{aug}}$ is calculated:
\begin{equation}
    \mathbf{V}_{\text{aug}} = \text{Concat}(\mathbf{V}, \mathbf{R}') \in \mathbb{R}^{(C + N) \times T \times H \times W},
\end{equation}
where $\mathbf{V}$ is the original video input with $C$ channels.
Convolutional filters in the 3D U-Net thereby operate on both local pixel data and broadcasted register-tokens simultaneously. The \textit{register-tokens} function as compact, learnable memory that injects global priors directly into the input. Each token is projected into a spatio-temporal map that covers all frames, allowing the encoder to assess local evidence and global context simultaneously in each layer. In effect, these maps serve as scene-dependent bias fields, directing the network to most noticeable and temporally stable locations. This method uses regular convolutions to induce attention-like effects and global coherence in the absence of an explicit attention module or recurrent training.

\paragraph{Encoder-Decoder}
In the saliency pretraining stage, the augmented input sequence $\mathbf{V}_{\text{aug}}$ (e.g. concatenation of the video and any auxiliary channels) is first processed by the UNet3D encoder, denoted as $\mathcal{E}$. The encoder extracts a spatio-temporal feature representation:
\begin{equation}
    \mathbf{Z} = \mathcal{E}(\mathbf{V}_{\text{aug}}),
\end{equation}
where $\mathcal{E}(\cdot)$ represents the complete stack of encoding layers.

At the bottleneck stage of the network, a spatial attention map is constructed to highlight salient regions. This is achieved by passing the encoder output $\mathbf{Z}$ through a 3D convolution and a sigmoid activation function, resulting in the mask $\mathcal{A}(\mathbf{Z})$:
\begin{equation}
    \mathcal{A}(\mathbf{Z}) = \sigma(\mathrm{Conv3D}(\mathbf{Z})),
\end{equation}
where $\sigma(\cdot)$ is the element-wise sigmoid function.

Next, the bottleneck features $\mathbf{Z}$ are further processed by a transformation module $\mathcal{B}$ (e.g., a 3D convolutional block), and modulated by the attention mask. $\mathbf{Z}$ are the output of the deepest encoder layer and represent a compact, highly abstract version of the input. The refined feature tensor $\mathbf{Z}'$ is therefore:
\begin{equation}
    \mathbf{Z}' = \mathcal{B}(\mathbf{Z}) \cdot \mathcal{A}(\mathbf{Z}),
\end{equation}
where $\cdot$ denotes element-wise multiplication. Finally, a decoder $\mathcal{D}$ reconstructs the predicted saliency sequence $\hat{\mathbf{S}}$ from the attended and transformed bottleneck features:
\begin{equation}
    \hat{\mathbf{S}} = \mathcal{D}(\mathbf{Z}'),
\end{equation}
with $\mathcal{D}(\cdot)$ representing the UNet3D decoder module. Our saliency backbone analyzes each video clip as a 3D volume and predicts saliency over space and time. The encoder combines appearance and motion cues across frames, using shared register-tokens and bottleneck attention to create a global prior. The decoder upsamples this representation to generate a saliency map for each frame in a single forward pass. This produces \textit{dynamic saliency}, which shifts with moving objects, while remaining temporally coherent.

\paragraph{Loss Functions}
To train the saliency prediction module, we employ two complementary loss functions that assess how well the predicted saliency map $\hat{\mathbf{S}}$ matches the ground truth saliency map $\mathbf{S}$ at the pixel level. Let $N$ denote the total number of pixels in each saliency map. For each pixel indexed by $j$ ($j = 1, 2, \ldots, N$), $\mathbf{S}_j$ and $\hat{\mathbf{S}}_j$ denote the ground-truth and predicted saliency values, respectively.

We define the mean value of the ground-truth saliency map as:
\begin{equation}
    \mu_{\mathbf{S}} = \frac{1}{N} \sum_{j=1}^N \mathbf{S}_j,
\end{equation}
and the mean value of the predicted map as:
\begin{equation}
    \mu_{\hat{\mathbf{S}}} = \frac{1}{N} \sum_{j=1}^N \hat{\mathbf{S}}_j.
\end{equation}

The KL divergence loss \cite{kullback1951information} measures the average distributional difference between the two maps on all pixels:
\begin{equation}
    \mathcal{L}_{\text{KL}} = D_{\text{KL}}(\mathbf{S} \parallel \hat{\mathbf{S}}) = \frac{1}{N} \sum_{j=1}^{N} \mathbf{S}_{j} \log \frac{\mathbf{S}_{j}}{\hat{\mathbf{S}}_{j}}.
\end{equation}

Additionally, the Pearson correlation coefficient loss \cite{benesty2009pearson} encourages linear correlation between predicted and ground-truth maps, and is computed as:
\begin{equation}
    \mathcal{L}_{\text{CC}} = - \frac{ \sum_{j=1}^{N} \left(\hat{\mathbf{S}}_{j} - \mu_{\hat{\mathbf{S}}}\right) \left(\mathbf{S}_{j} - \mu_{\mathbf{S}}\right) }
    { \sqrt{ \sum_{j=1}^{N} \left(\hat{\mathbf{S}}_{j} - \mu_{\hat{\mathbf{S}}}\right)^2 } \sqrt{ \sum_{j=1}^{N} \left(\mathbf{S}_{j} - \mu_{\mathbf{S}}\right)^2 } }.
\end{equation}
All summations and averages are computed over the $N$ pixels indexed by $j$ in each map.

\begin{figure*}[ht]
\centering
\includegraphics[width=1.0\textwidth]{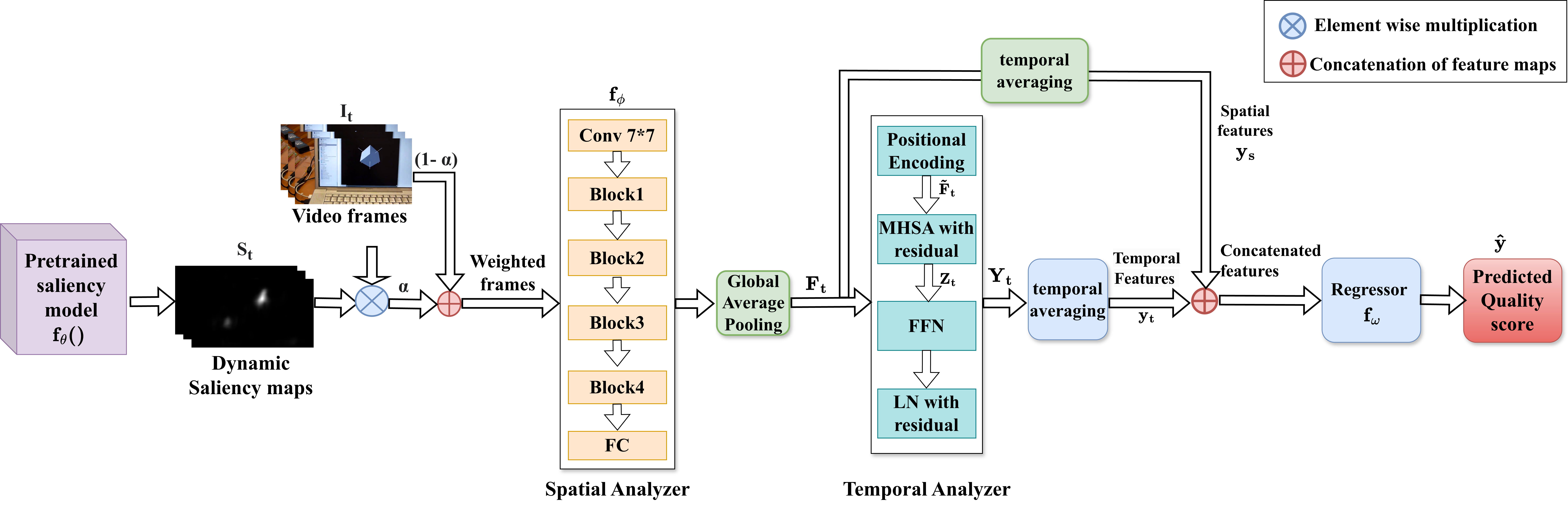}
\caption{Overview of the proposed spatio-temporal VQA module. A pretrained saliency model $f_{\theta}()$ produces dynamic saliency maps to form saliency-weighted video frames, which pass through a ResNet-50 backbone to extract and pool spatial features $F_{t}$. These are temporally pooled to $y_{s}$. Then, processed by a Transformer encoder, the sequence is concatenated and projected with pooled temporal features $y_{t}$ before being fed to a regressor $f_{\omega}$, which predicts the video quality score $\hat{y}$. LN denotes layer normalization, FFN is Feed Forward Network, MHSA is multi-head self-attention, Conv is the initial convolution, Block~1–4 means the ResNet convolution stages, and FC means a fully connected layer.}
\label{fig:spatiotemporal_vqa}
\end{figure*}

Finally, we combine both objectives into a single weighted loss for training:
\begin{equation}
    \mathcal{L} = \gamma \cdot \mathcal{L}_{\text{KL}} + \mathcal{L}_{\text{CC}},
\end{equation}
where $\gamma$ is a hyperparameter that controls the trade-off between the two loss terms.

\subsection{Spatio-temporal Video Quality Analyzer}

Figure~\ref{fig:spatiotemporal_vqa} shows the Spatio-Temporal Video Quality Analyzer module that integrates saliency-driven spatial encoding and transformer-based temporal modeling to produce perceptually relevant and temporally coherent video quality predictions.   For an input video  $\mathbf{V}$, the frame at a specific time stamp \( t \) is represented as $\mathbf{I}_t$. We derive  dynamic saliency maps of the frames $\mathbf{I}_t$, denoted as $\mathbf{S}_t$, using the model described in Section \ref{fig:saliency_pretraining}, as given by:
\begin{equation}
\mathbf{S}_t = f_{\theta}(\mathbf{I}_t), \quad \mathbf{S}_t \in \mathbb{R}^{H \times W},
\end{equation}
where $f{\theta}(\cdot)$ is the saliency network.

For each frame, we construct a fused input by combining the RGB image $\mathbf{I}_t$ with its predicted saliency map $\mathbf{S}_t$ using a weighted sum. The spatial features $\mathbf{F}_t$ are then extracted and pooled from the fused inputs using a ResNet-50 encoder, as follows.
\begin{equation}
\mathbf{F}_t = \text{GAP}\!\left(
    f_{\phi}\big( (1 - \alpha)\mathbf{I}_t + \alpha (\mathbf{I}_t \odot \mathbf{S}_t) \big)
\right),
\end{equation}
where $f_\phi$ denotes the ResNet-50 encoder, $\text{GAP}(\cdot)$ denotes global average pooling over the spatial dimensions, $\mathbf{I}_t$ is the original frame, $\mathbf{S}_t$ is the saliency map, $\odot$ is element-wise multiplication and $\alpha$ controls fusion. For each frame, $\alpha$ sets the proportion of saliency-enhanced information versus original features in the final representation.

Next, Sine-cosine Position Encodings (PE) are added to spatial features $\mathbf{F}_t$ to provide temporal order, as follows.
\begin{equation}
\tilde{\mathbf{F}}_t = \mathbf{F}_t + \mathbf{P}_t, \qquad \mathbf{P}_t = \text{PE}(t),
\end{equation}
 where $\mathbf{F}_t$ is the fused spatial feature of frame $t$, and $\text{PE}(t)$ denotes the positional encoding at time $t$. To model temporal dependencies across frames, we process each position-enhanced feature vector $\tilde{\mathbf{F}}_t$ for frame $t$ with a lightweight temporal transformer encoder. The transformer is composed of two encoder layers, each with multi-head self-attention and feed-forward network sub-blocks as well as skip (residual) connections and layer normalization.

In each encoder layer, $\mathbf{Z}_t$ is computed as the attended, contextually updated representation for frame $t$, with a residual connection added to the input, as follows. 
\begin{align}
\mathbf{Z}_t &= \text{MHSA}(\tilde{\mathbf{F}}_t) + \tilde{\mathbf{F}}_t ,
\end{align}
where $\text{MHSA}(\cdot)$ is multi-head self-attention operation that is applied to the input feature $\tilde{\mathbf{F}}_t$.

Next, this output $\mathbf{Z}_t$ passes through a feed-forward network, again augmented with a residual connection and followed by layer normalization to represent $\mathbf{Y}_t$:
\begin{align}
\mathbf{Y}_t &= \text{LN}(\text{FFN}(\mathbf{Z}_t) + \mathbf{Z}_t) ,
\end{align}
where $\text{FFN}(\cdot)$ is a feed-forward neural network and $\text{LN}(\cdot)$ denotes normalization layer.

We obtain a mean spatial feature $\mathbf{y}_{\text{s}}$ by first applying global average pooling over the spatial dimensions of each frame feature map $\mathbf{F}_t$ and then averaging these vectors over time. In parallel, we obtain a mean temporal feature $\mathbf{y}_{\text{t}}$ by averaging the transformer outputs $\mathbf{Y}_t$ over time. These are concatenated and passed through a fully connected regression layer, which maps this feature representation to the final video quality prediction $\hat{y}$ as follows:
\begin{equation}
\hat{y} = f_{\omega}\big([\mathbf{y}_{\text{s}};\mathbf{y}_{\text{t}}]\big).
\end{equation}

The VQA training employs a joint loss for MOS regression, which is different from the saliency pretraining losses. More specifically, for MOS prediction, we use the following weighted Loss:
\begin{equation}
\mathcal{L} = \mathcal{L}_{\text{L1}} + \beta \cdot \mathcal{L}{\text{corr}} ,
\end{equation}
where $\mathcal{L}_{\text{L1}}$ is the absolute error loss $|\hat{y} - y|$, $\mathcal{L}_{\text{corr}}$ is the Spearman rank correlation loss $(1 - \rho(\hat{y}, y))$ \cite{spearman1904general}, and $\beta$ is a fixed loss weight. 

This setup ensures that the perceptual quality is consistently modeled and robust to the variation of the MOS specific to the dataset. By combining saliency-enhanced spatial features with transformer-based temporal modeling and robust training objectives, our model produces perceptually aligned and temporally consistent quality estimates across diverse datasets.

\section{Computational Complexity}
The proposed DAGR-VQA model is designed to process video sequences with high efficiency through the integration of convolutional feature extraction and temporal attention mechanisms. To evaluate its computational cost, we conduct an analysis of the complexity inherent in its core components and compare it with transformer-based and sampling-based methodologies as follows.
\begin{itemize}
    \item {Saliency Prediction (UNet3D)}:
 Hierarchical convolution operations have a complexity of $O(T \cdot N \cdot d)$, where \(T\) is the number of frames, \(N = H \times W\) represents spatial tokens per frame after resizing to \(224\times398\), and \(d\) is the dimension of the characteristic. Register-tokens are incorporated via single Conv3D operations, adding negligible overhead.
\item {Spatial Feature Extraction (ResNet50)}: 
Processing saliency-weighted frames has a complexity of $O(T \cdot N \cdot d)$.
\item {Temporal Modeling (Transformer):}  
The temporal transformer introduces self-attention over frame sequences with a complexity of $O(T^2 \cdot d)$.
\item {Overall Complexity:}  
The total computational cost is given by $O(T \cdot N \cdot d + T^2 \cdot d)$. For our settings (\(T=8\), \(N \gg T\)), spatial processing dominates, producing an effective complexity of \(O(T \cdot N \cdot d)\).
\end{itemize}
In conclusion, the ViViT model employs self-attention mechanisms across all spatio-temporal tokens, resulting in a complexity of $O((T \cdot N)^2 \cdot d)\). For an input dimension of \(8\times224\times224\), this approach necessitates approximately {141 GFLOPs}. In contrast, the DAGR-VQA model requires {59 GFLOPs}, thereby achieving a reduction in computational demand by a factor of 2.4. FAST-VQA~\cite{wu2022fastvqa} attains computational efficiency through the implementation of Grid Mini-patch Sampling, utilizing \(G_f^2\) fragments per frame. The primary computational burden of its transformer, when processing across \(T\) frames, follows a complexity of 
$O\bigl(T \times G_f^2 \times d\bigr)$. Under standard configuration parameters where \(G_f=7\) and \(T=8\), the computational workload amounts to {279 GFLOPs}. In contrast, the lightweight variant FAST-VQA-M, characterized by \(G_f=4\)) requires  {46 GFLOPs}, a reduction achieved through significant downsampling. In contrast, DAGR-VQA necessitates {59 GFLOPs}, thereby providing an equilibrium between computational efficiency and comprehensive spatio-temporal sequence processing.

\begin{table*}[tb]
    \centering
    \caption{Performance comparison of our method with state-of-the-art NR-VQA models across four datasets. The best three results for each metric are in red, green and blue, respectively.}
    \begin{tabular}{lcccccccccc}
        \toprule
        \multirow{2}{*}{\textbf{Method}} & \multicolumn{2}{c}{\textbf{LSVQ}} & \multicolumn{2}{c}{\textbf{KonVid-1k}} & \multicolumn{2}{c}{\textbf{LIVE-VQC}} & \multicolumn{2}{c}{\textbf{YouTube-UGC}} & \multicolumn{2}{c}{\textbf{Average}} \\
        & \textbf{PLCC} & \textbf{SRCC} & \textbf{PLCC} & \textbf{SRCC} & \textbf{PLCC} & \textbf{SRCC} & \textbf{PLCC} & \textbf{SRCC} & \textbf{PLCC} & \textbf{SRCC} \\
        \midrule
        BRISQUE \cite{mittal2011brisque} & 0.576 & 0.579 & 0.658 & 0.657 & 0.638 & 0.593 & 0.395 & 0.382 & 0.567 & 0.553 \\
        VIDEVAL \cite{tu2021videval} & 0.783 & 0.794 & 0.780 & 0.783 & 0.751 & 0.752 & 0.773 & 0.779 & 0.772 & 0.777 \\
        StarVQA \cite{xing2022starvqa} & 0.857 & 0.851 & 0.796 & 0.812 & 0.808 & 0.732 & N/A & N/A & 0.820 & 0.798 \\
        Patch-VQ \cite{ying2021patchvq} & 0.828 & 0.827 & 0.786 & 0.791 & 0.837 & 0.827 & N/A & N/A & 0.817 & 0.815 \\
        FastVQA \cite{wu2022fastvqa} & 0.877 & 0.876 & 0.855 & 0.859 & 0.844 & 0.823 & 0.852 & 0.855 & 0.857 & 0.853 \\
        MaxVQA \cite{wu2023maxvqa} & N/A & N/A & \color{blue}{0.895} & 0.894 & 0.873 & 0.854 & 0.890 & \color{blue}{0.894} & {0.886} & {0.881} \\
        PTM-VQA \cite{yuan2024ptmvqa} & 0.864 & 0.855 & 0.872 & 0.857 & 0.820 & 0.811 & 0.857 & 0.858 & 0.853 & 0.845 \\
        DOVER \cite{Wu2023Dover} & 0.889 & 0.888 & \color{green}{0.906} & \color{green}{0.909} & 0.875 & 0.860 & 0.891 & 0.890 & \color{blue}{0.890} & \color{blue}{0.888} \\
        FineVQ \cite{duan2025finevq} & \color{red}{0.900} & \color{green}{0.900} & \color{red}{0.915} & \color{red}{0.910} & \color{blue}{0.895} & \color{red}{0.895} & \color{green}{0.910} & \color{red}{0.914} & \color{red}{0.905} & \color{red}{0.905} \\
        KVQ \cite{qu2025kvq} & \color{green}{0.897} & \color{blue}{0.896} & {0.892} & 0.890 & 0.843 & 0.820 & N/A & N/A & 0.877 & 0.868 \\
        UNISAL~\cite{droste2020unified} + SpatiotempVQA   & 0.839 & 0.872 & 0.889 & 0.859 & 0.891 & 0.872 & 0.876 & 0.875 & 0.874 & 0.870 \\
        DiffSal~\cite{xiong2024diffsal} + SpatiotempVQA   & 0.847 & 0.879 & 0.885 & 0.867 & 0.893 & 0.874 & 0.883 & 0.884 & 0.877 & 0.876 \\
        ViNet-S~\cite{girmaji2025vinet} + SpatiotempVQA   & 0.850 & 0.885 & 0.887 & 0.872 & \color{green}{0.897} & \color{blue}{0.883} & \color{blue}{0.890} & 0.887 & 0.881 & 0.882 \\
        \textbf{Ours} & \color{blue}{0.892} & \color{red}{0.907} & 0.863 & \color{blue}{0.896} & \color{red}{0.915} & \color{green}{0.886} & \color{red}{0.913} & \color{green}{0.910} & \color{green}{0.896} & \color{green}{0.900} \\
        \bottomrule
    \end{tabular}
    \label{tab:comparison}
\end{table*}

\section{Experimental Results}

We evaluated DAGR-VQA on four standard no-reference VQA datasets: LSVQ~\cite{ying2021patchvq}, KonVid-1k~\cite{hosu2017konstanz}, LIVE-VQC~\cite{sinno2018large}, and YouTube-UGC~\cite{tu2020ugc}. These datasets feature diverse real-world videos with varying distortions, resolutions, and durations, providing a comprehensive benchmark for robustness. 

Our experimental pipeline consists of saliency pre-training followed by spatio-temporal feature extraction for VQA. The saliency model is pre-trained on the DHF1K eye-tracking dataset~\cite{wang2018revisiting} for 180 epochs using the Adam optimizer (learning rate $5 \times 10^{-3}$, batch size 4, loss weight $\gamma = 0.01$). The VQA model is trained for 300 epochs with Adam (learning rate $1 \times 10^{-5}$, batch size 5) and Cosine Annealing scheduler. We set the number of register-tokens to $N = 4$ and use a saliency weighting factor $\alpha = 0.5$ for spatial fusion. The training objective combines $L_1$ regression loss and Spearman correlation, with the ranking loss weighted by $\beta = 0.1$. Both modules use an 80:10:10 split for train, validation, and test sets. Model performance is evaluated using Spearman Rank Correlation Coefficient (SRCC) and Pearson Linear Correlation Coefficient (PLCC).

To optimize computational efficiency while preserving content fidelity, we applied uniform frame sampling with linear interpolation, following recommendations from recent works~\cite{sun2024analysis, hu2024video}. For saliency prediction, 60 frames per video are sampled to capture detailed spatio-temporal dynamics. For VQA, only 8 equidistant frames are selected for efficiency. The sampling involved selecting $N$ evenly spaced frames from the original video sequence of length $L$. To be more precise, the algorithm divides the video into $N$ segments and selects one representative frame from each segment to ensure uniform coverage across the entire sequence. All frames are resized to $224 \times 398$ pixels.

Table~\ref{tab:comparison} summarizes the performance of our approach alongside a comprehensive set of NR-VQA baselines across four benchmark datasets. The baselines include traditional feature-based methods, and a diverse selection of state-of-the-art deep learning models, such as Patch-VQ~\cite{ying2021patchvq}, FastVQA~\cite{wu2022fastvqa}, KVQ~\cite{qu2025kvq}. We also report results for variants of our framework that leverage leading pre-trained saliency models as plug-in backbones (e.g. UNISAL~\cite{droste2020unified}) integrated via the spatio-temporal VQA pipeline. Our method consistently achieves highly competitive, and often superior, performance across all datasets. On LSVQ, it yields the highest SRCC (0.907), outperforming FineVQ and KVQ, and ranks third in PLCC (0.892). For KonVid-1k, it achieves the second-best SRCC (0.896), surpassing MaxVQA, with a PLCC (0.863) slightly below the top three. On LIVE-VQC, our approach delivers the best PLCC (0.915) and the second-best SRCC (0.886), outperforming MaxVQA and closely following FineVQ. On YouTube-UGC, it again leads in PLCC (0.913) and ranks second in SRCC (0.910), demonstrating robust performance on user-generated content. Averaged across all datasets, our approach achieves a PLCC of 0.896 and an SRCC of 0.900, ranking second only to FineVQ and outperforming MaxVQA and other strong baselines.

These findings underscore the robustness of our method, where it often exceeds or closely tracks the leading approaches. On LIVE-VQC, DAGR-VQA outperforms FineVQ (SRCC of 0.886 vs 0.895), whereas on KonVid-1k, FineVQ yields higher correlations (SRCC of 0.910 vs 0.896). To rigorously assess the significance of these differences, we performed both paired t-tests and Wilcoxon signed-rank tests on the per-video predictions for each dataset in Table \ref{tab:stat-results}. In all cases, the $p$-values are well above the standard significance threshold of $0.05$, indicating that the observed differences between DAGR-VQA and FineVQ are not statistically significant. In addition to comparable predictive accuracy, DAGR-VQA offers faster inference and higher throughput (FPS) than FineVQ (see Table~\ref{tab:inference_runtime_1080p}), making it well-suited for real-time and large-scale scenarios.

\begin{table}[h]
    \centering
    \caption{p-values comparing DAGR-VQA and FineVQ.}
    \begin{tabular}{lcc}
        \toprule
        Test & LIVE-VQC & KonVid-1k \\
        \midrule
        Paired t-test          & 0.41  & 0.094 \\
        Wilcoxon signed-rank   & 0.39  & 0.088 \\
        \bottomrule
    \end{tabular}
    \label{tab:stat-results}
\end{table}

\begin{figure}[tb]
    \centering 
    \includegraphics[width=0.45\textwidth]{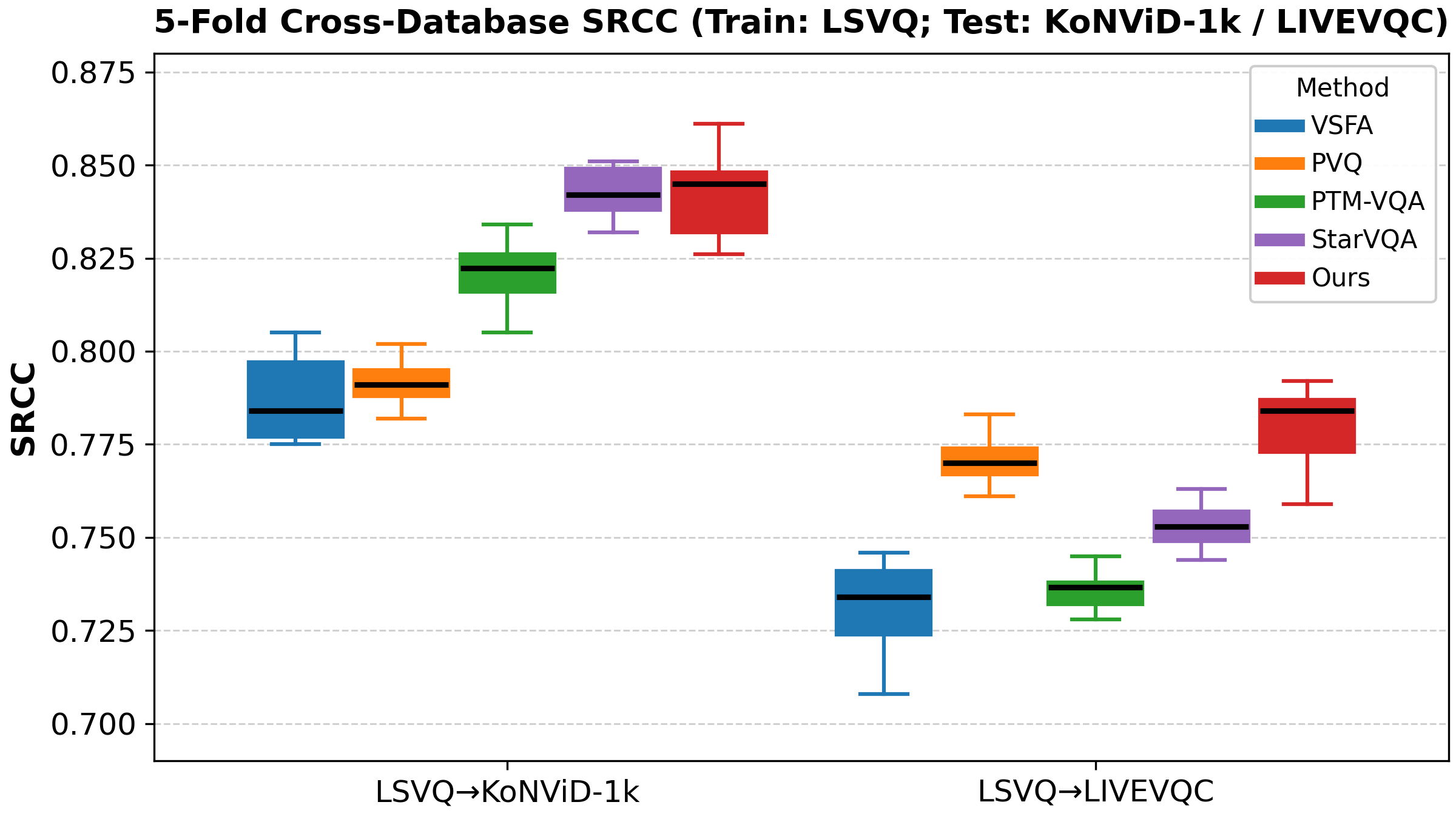}  
    \caption{5-Fold Cross-Database SRCC (Train: LSVQ $\rightarrow$ Test: KoNViD-1k / LIVEVQC). Each box plot shows results for a VQA method, where the central black line within each box indicates the median SRCC across 5 runs.}
    \label{fig:crossval}
\end{figure}

To thoroughly evaluate our model’s generalization capability among different datasets, we conducted a 5-fold cross-dataset experiment in which LSVQ was used for training, while KoNViD-1k and LIVE-VQC served as distinct test sets. Figure~\ref{fig:crossval} illustrates the SRCC distributions obtained across five folds for each train/test configuration in y-axis, with the \textit{x}-axis indicating the respective dataset pairs. Each box plot displays the performance of a specific method, with the central black line marking the median SRCC for the five runs. We compared DAGR-VQA to four established no-reference VQA baselines: VSFA~\cite{li2019quality}, patch-based PVQ~\cite{ying2021patchvq}, PTM-VQA~\cite{yuan2024ptmvqa}, and StarVQA~\cite{xing2022starvqa}. In both cross-database evaluations (LSVQ$\rightarrow$KoNViD-1k and LSVQ$\rightarrow$LIVE-VQC), our model attained the highest median and overall SRCC, outperforming all alternative approaches. This indicates that DAGR-VQA not only achieves superior performance on individual datasets but also exhibits robust generalization to novel video content and previously unencountered distortion types.

To evaluate the effectiveness of register token augmentation within our saliency framework, we benchmarked the model on both the DHF1K and DIEM validation sets, using three established saliency metrics: Normalized Scanpath Saliency (NSS)~\cite{Judd09}, which quantifies correspondence between predicted saliency maps and human fixation locations; Correlation Coefficient (CC)~\cite{Bruce15}, assessing the linear relationship between predicted and ground-truth maps; and AUC-Judd (AUC-J)~\cite{Borji13}, which measures the area under the ROC curve for predicting fixations. As shown in Table~\ref{tab:saliency_eval}, our full UNet3D+RT model significantly outperforms recent state-of-the-art methods, including ViNet-S, TASED-Net, and our own baseline without register-tokens. On DHF1K, our model achieves an NSS of 3.683, CC of 0.704, and AUC-J of 0.942; for DIEM, the respective numbers are 2.856, 0.685, and 0.902, setting a new performance standard on both datasets. Notably, the improvement over UNet3D without RT and other competitors is consistent across all three metrics, underscoring the robust benefit of incorporating global register-tokens. These results verify that adding register-tokens enables the network to integrate scene-wide, long-range information, producing saliency predictions with markedly higher alignment to human fixations in dynamic settings.

\begin{table*}[ht]
\centering
\caption{Quantitative evaluation of saliency prediction models on DHF1K and DIEM validation sets. The best entries are highlighted in bold.}
\setlength{\tabcolsep}{8pt}
\resizebox{\textwidth}{!}{%
\begin{tabular}{l cc cc cc cc cc cc cc cc cc}
\toprule
\textbf{Metric} 
    & \multicolumn{2}{c}{ACLNet} 
    & \multicolumn{2}{c}{ViNet-S}
    & \multicolumn{2}{c}{TASED-Net}
    & \multicolumn{2}{c}{DeepVS}
    & \multicolumn{2}{c}{DiffSal}
    & \multicolumn{2}{c}{UNet3D (Ours, w/o RT)}
    & \multicolumn{2}{c}{\textbf{UNet3D+RT (Ours)}} \\
\cmidrule(lr){2-3}  \cmidrule(lr){4-5}  \cmidrule(lr){6-7} \cmidrule(lr){8-9} 
\cmidrule(lr){10-11} \cmidrule(lr){12-13} \cmidrule(lr){14-15} \cmidrule(lr){16-17} \cmidrule(lr){18-19}
& DHF1K & DIEM
& DHF1K & DIEM
& DHF1K & DIEM
& DHF1K & DIEM
& DHF1K & DIEM
& DHF1K & DIEM
& DHF1K & DIEM
\\ \midrule
NSS   & 2.35 & 2.02 & 3.008 & 2.732 & 2.706 & 2.16 & 2.774 & 2.25 & 3.066 & 2.65 & 2.945 & 2.720 & \textbf{3.683} & \textbf{2.856} \\
CC    & 0.434 & 0.522 & 0.529 & 0.673 & 0.481 & 0.557 & 0.525 & 0.49 & 0.533 & 0.66 & 0.640 & 0.675 & \textbf{0.704} & \textbf{0.685} \\
AUC-J & 0.890 & 0.869 & 0.919 & \textbf{0.908} & 0.894 & 0.881 & 0.887 & 0.86 & 0.918 & 0.907 & 0.904 & 0.897 & \textbf{0.942} & 0.902 \\
\bottomrule
\end{tabular}
}
\label{tab:saliency_eval}
\end{table*}

\begin{table*}[!ht]
\centering
\caption{Inference time and FPS for 10 videos ($240$ frames each) at 1080p. The best entries are highlighted in bold.}
\label{tab:inference_runtime_1080p}
\resizebox{0.95\textwidth}{!}{%
\begin{tabular}{lcccccccc}
\toprule
\textbf{Metric} &
TLVQM\cite{korhonen2019tlvqm} & VIDEVAL\cite{tu2021videval} & VSFA\cite{li2019quality} & SimpleVQA\cite{sun2022simplevqa} & FastVQA\cite{wu2022fastvqa} & Dover\cite{Wu2023Dover} & FineVQ\cite{duan2025finevq} & DAGR-VQA \\
\midrule
\textbf{Inference Time (s)} 
   & 476.5    & 1059.2   & 53.6    & 40.3    & \textbf{7.39}   & \textbf{5.07}   & 10.46   & \textbf{6.19}   \\
\textbf{FPS} 
   & 5.04     & 2.27     & 44.80   & 59.6    & \textbf{324.7}  & \textbf{473.4}  & 229.4   & \textbf{387.7}  \\
\bottomrule
\end{tabular}
}
\end{table*}

We compared the computational efficiency of DAGR-VQA with state-of-the-art VQA methods on ten 1080p videos (1920x1080), each with 240 frames. To ensure consistency with the reporting standards of the community~\cite{duan2025finevq, Wu2023Dover, wu2022fastvqa}, frames-per-second (FPS) values are normalized based on the total number of test frames (specifically, $10 \times 240 = 2400$), irrespective of the actual sampling of frames. Inference time is measured directly on an RTX A5000 GPU with mixed precision enabled. As shown in Table~\ref{tab:inference_runtime_1080p}, DAGR-VQA achieves a normalized FPS of $387.7$ at 1080p, which is fairly competitive with recent VQA benchmarks. All FPS values are normalized to the total available video frames for direct comparison with previous work~\cite{duan2025finevq, Wu2023Dover, wu2022fastvqa}. In addition to its strong system-level efficiency (see Table~\ref{tab:inference_runtime_1080p}), DAGR-VQA achieves the second-best overall accuracy on four main NR-VQA benchmarks, with an average SRCC of \textbf{0.900} and PLCC of \textbf{0.896} (see Table~\ref{tab:comparison}). 


\begin{figure}[ht]
    \centering
    \includegraphics[width=1.0\linewidth]{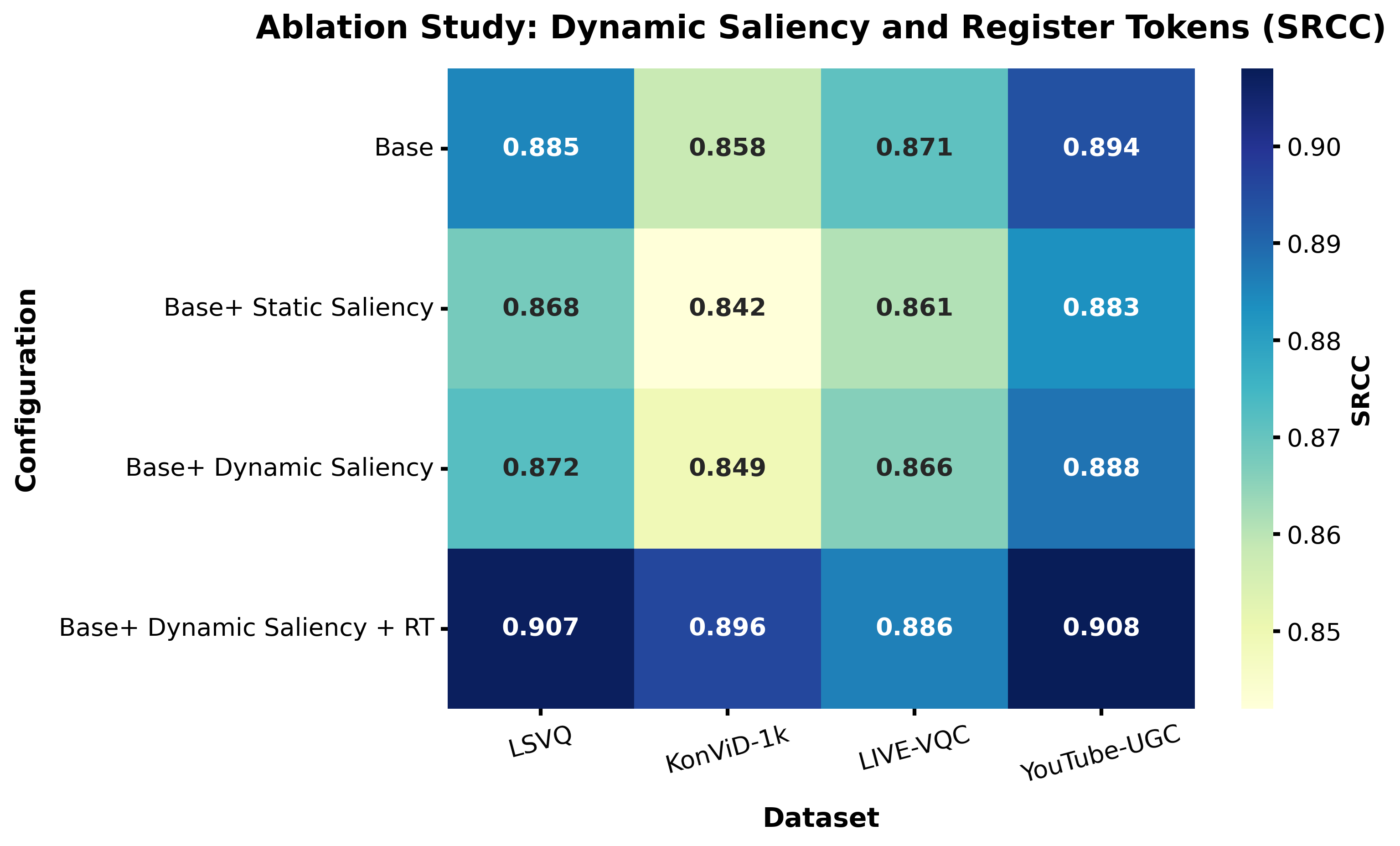} 
    \caption{
        Ablation study: SRCC of configurations with/without dynamic/static saliency and register-tokens across four datasets. Higher is better.
    }
    \label{fig:saliency-rt-heatmap}
\end{figure}

\section{Ablation Study}

To better understand the individual and joint contributions of dynamic saliency modeling and register-tokens to overall VQA accuracy, we performed an ablation study across the four benchmark datasets. Figure~\ref{fig:saliency-rt-heatmap} visually summarizes these results as a heatmap of SRCC values for all major model variants. The results indicate that the base model provides strong baseline performance, while adding static or dynamic saliency alone yields comparable or slightly lower accuracy, likely due to imperfect attention focusing in some video sequences. However, introducing register-tokens in conjunction with dynamic saliency consistently leads to substantial SRCC gains across all datasets, culminating in the highest performance on LSVQ (0.907), KonViD-1k (0.896), LIVE-VQC (0.886), and YouTube-UGC (0.908). These findings underscore that register-tokens robustly enhance the stability and effectiveness of dynamic saliency integration, enabling the model to better attend to content-relevant regions and maintain strong generalization.



Subsequently, Table~\ref{tab:dynsal_lsvq_konvid} provides a qualitative analysis featuring saliency map images produced by the proposed model for specific frames from the LSVQ and KonViD-1k datasets. These illustrations show how the dynamic saliency prediction with register-tokens preserves attention that is temporally coherent and sensitive to varying scene contexts. For example, within a video sequence derived from the LSVQ dataset, the images show that the model reliably tracks the most salient movements occurring throughout a basketball game, with the focus transitioning from the ``ball handler'' to ``on-court dynamics'', and culminating at the most salient moment at the rim, thereby showcasing seamless and substantive saliency transitions. Within the KonViD-1k example, register-tokens effectively direct the model attention across a rotating cube, emphasizing high-contrast and motion-relevant areas such as edges, vertices, and illuminated surfaces, while also incorporating secondary visual distractions. Compared to a baseline without register-tokens, these \textit{visual} findings demonstrate enhanced spatial stability and temporal consistency. Consequently, the dynamic saliency prediction adapts seamlessly to object movement and scene semantics, closely resembling human visual attention.

\begin{table*}[t]
\centering
\caption{
  \textbf{Dynamic saliency with register-tokens for LSVQ and KonViD-1k samples.}
  Each triplet shows the original frame, its overlayed dynamic saliency heatmap (with RT), and a concise narrative. The model’s attention adaptively shifts in time, following action and scene context, illustrating true HVS-inspired temporal reactivity.
}
\vspace{1pt}
\setlength{\tabcolsep}{1.0pt} 
\renewcommand{\arraystretch}{0.7} 
\resizebox{0.7\textwidth}{!}{%
\begin{tabular}{| >{\tiny\raggedright\arraybackslash}m{1.6cm} | >{\tiny\raggedright\arraybackslash}m{1.6cm} | >{\tiny\raggedright\arraybackslash}m{4.0cm} |}
\hline
\multicolumn{3}{|c|}{\tiny \textbf{LSVQ: Dynamic Saliency with Register-Tokens}} \\ \hline
\textbf{Original Frame} & \textbf{Saliency Overlay} & \textbf{Narrative} \\ \hline
\includegraphics[width=1.5cm]{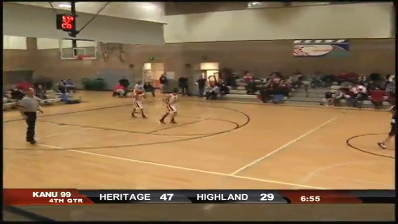} &
\includegraphics[width=1.5cm]{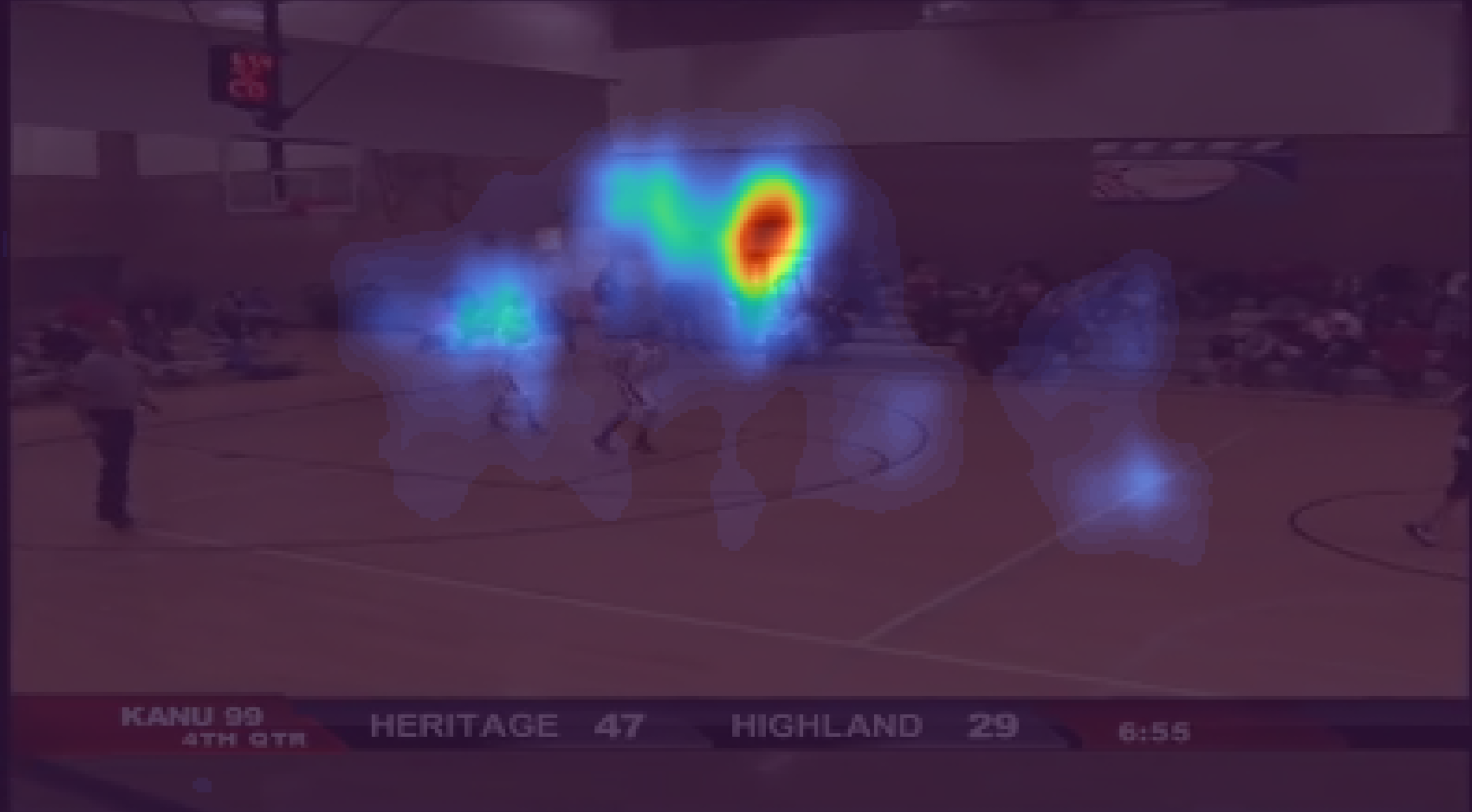} &
\tiny{Saliency focuses on ``white jersey players initiating play'' and on ``the ball handler'''
 as the team advances from the backcourt.} \\ \hline
\includegraphics[width=1.5cm]{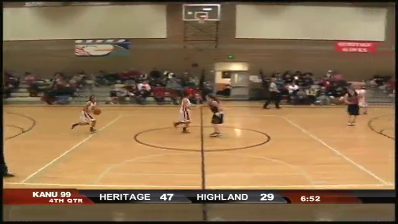} &
\includegraphics[width=1.5cm]{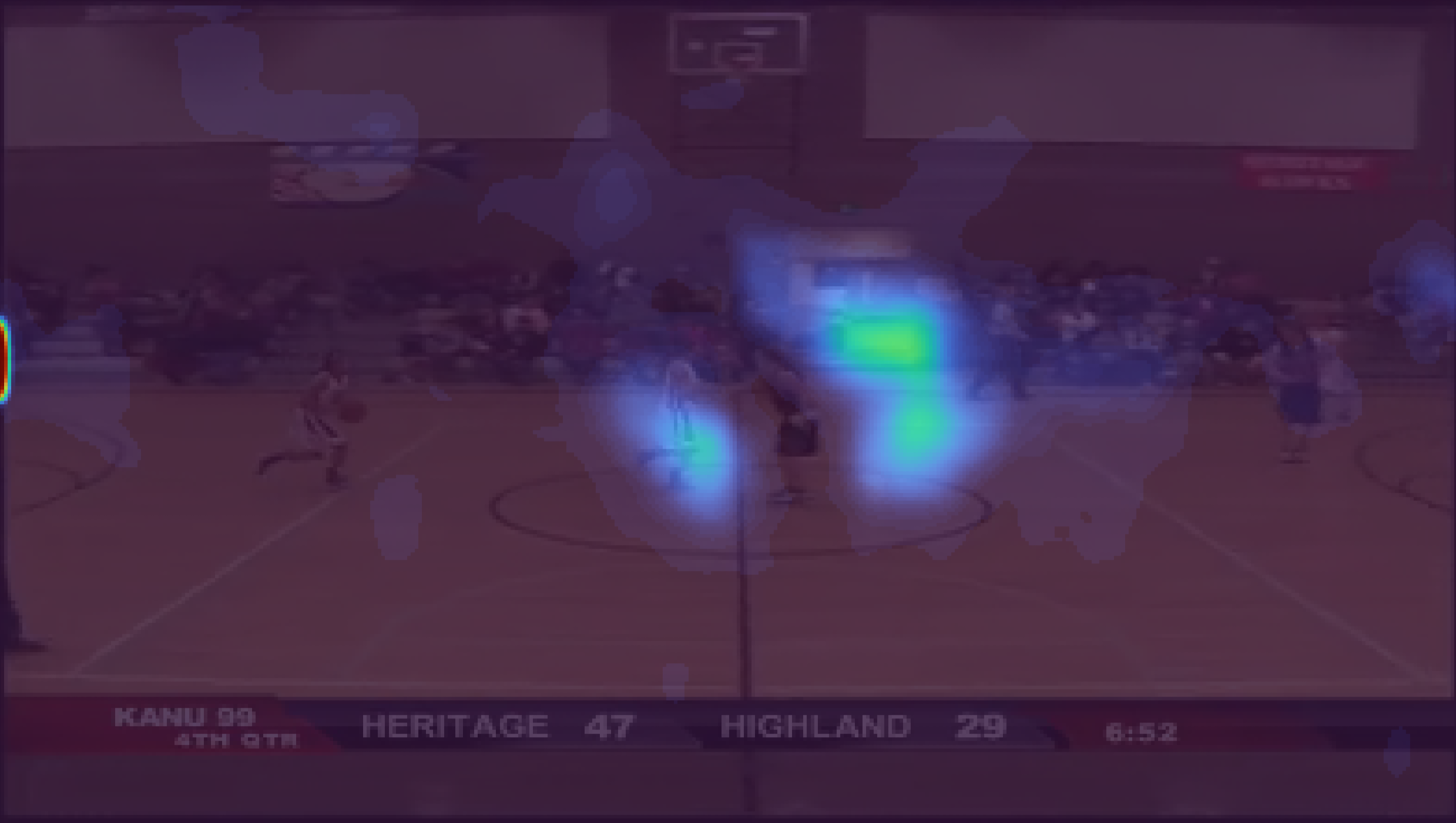} &
\tiny{Attention follows the ``ball-handler crossing midcourt'', flexibly tracking the offensive move and adapting to player and ball position changes.} \\ \hline
\includegraphics[width=1.5cm]{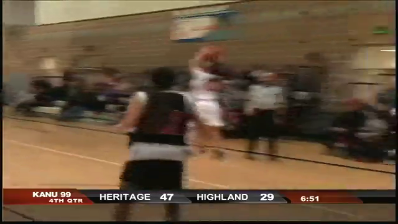} &
\includegraphics[width=1.5cm]{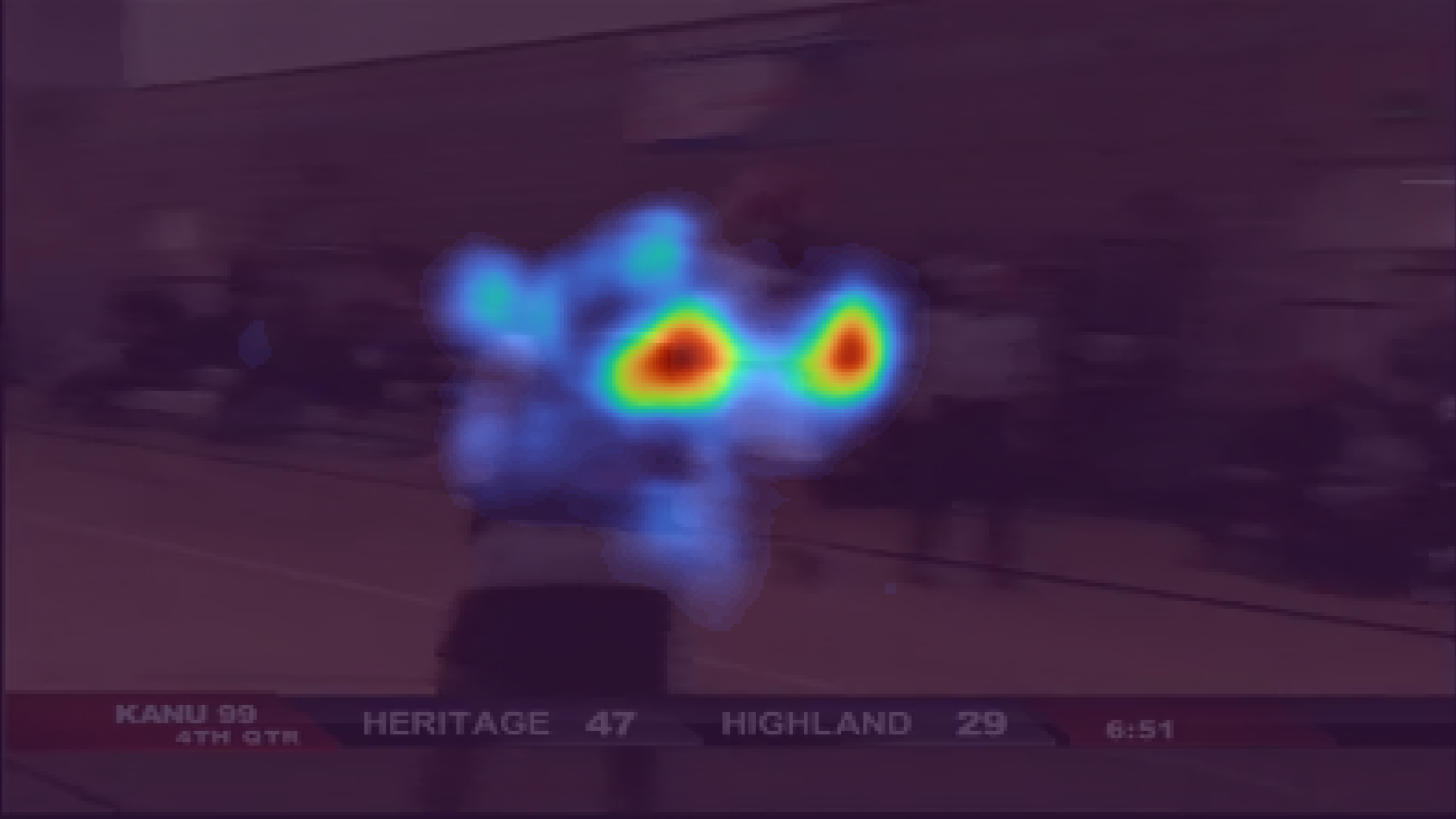} &
\tiny{Saliency centers on ``black jersey players'' during the climax of play, highlighting defensive actions and the importance of their positioning.} \\ \hline
\multicolumn{3}{|c|}{\tiny \textbf{KonViD-1k: Dynamic Saliency with Register-Tokens}} \\ \hline
\textbf{Original Frame} & \textbf{Saliency Overlay} & \textbf{Narrative} \\ \hline
\includegraphics[width=1.5cm]{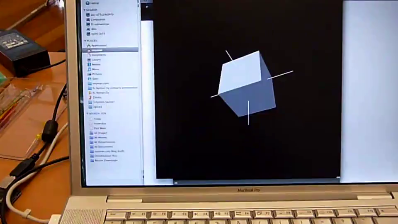} &
\includegraphics[width=1.5cm]{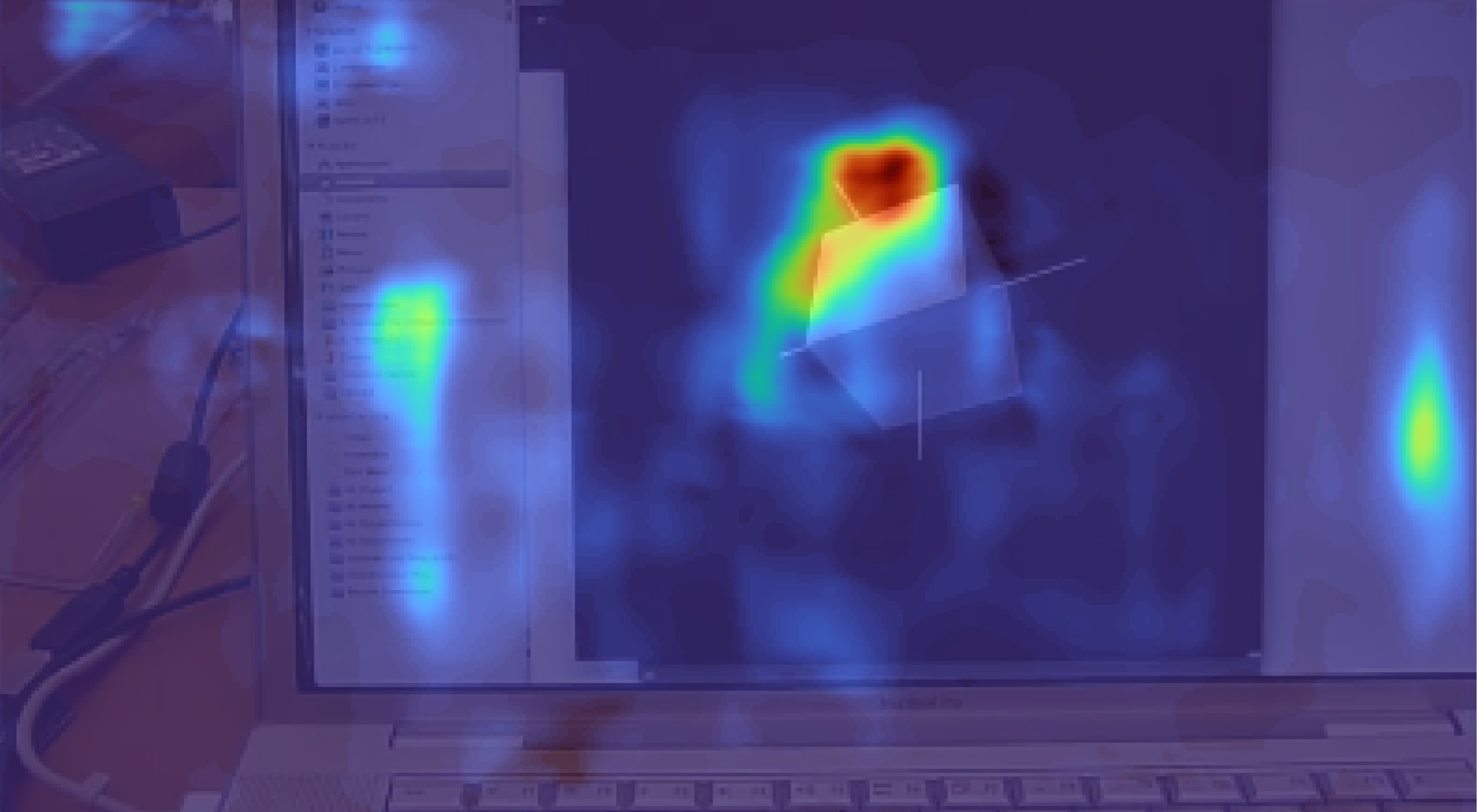} &
\tiny{As the cube rotates, attention is drawn to its bright top edge, which stands out as the most visually distinct moving part.} \\ \hline
\includegraphics[width=1.5cm]{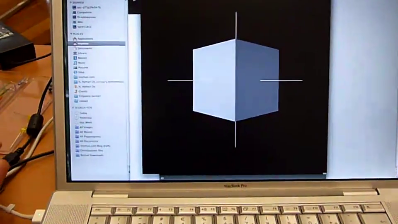} &
\includegraphics[width=1.5cm]{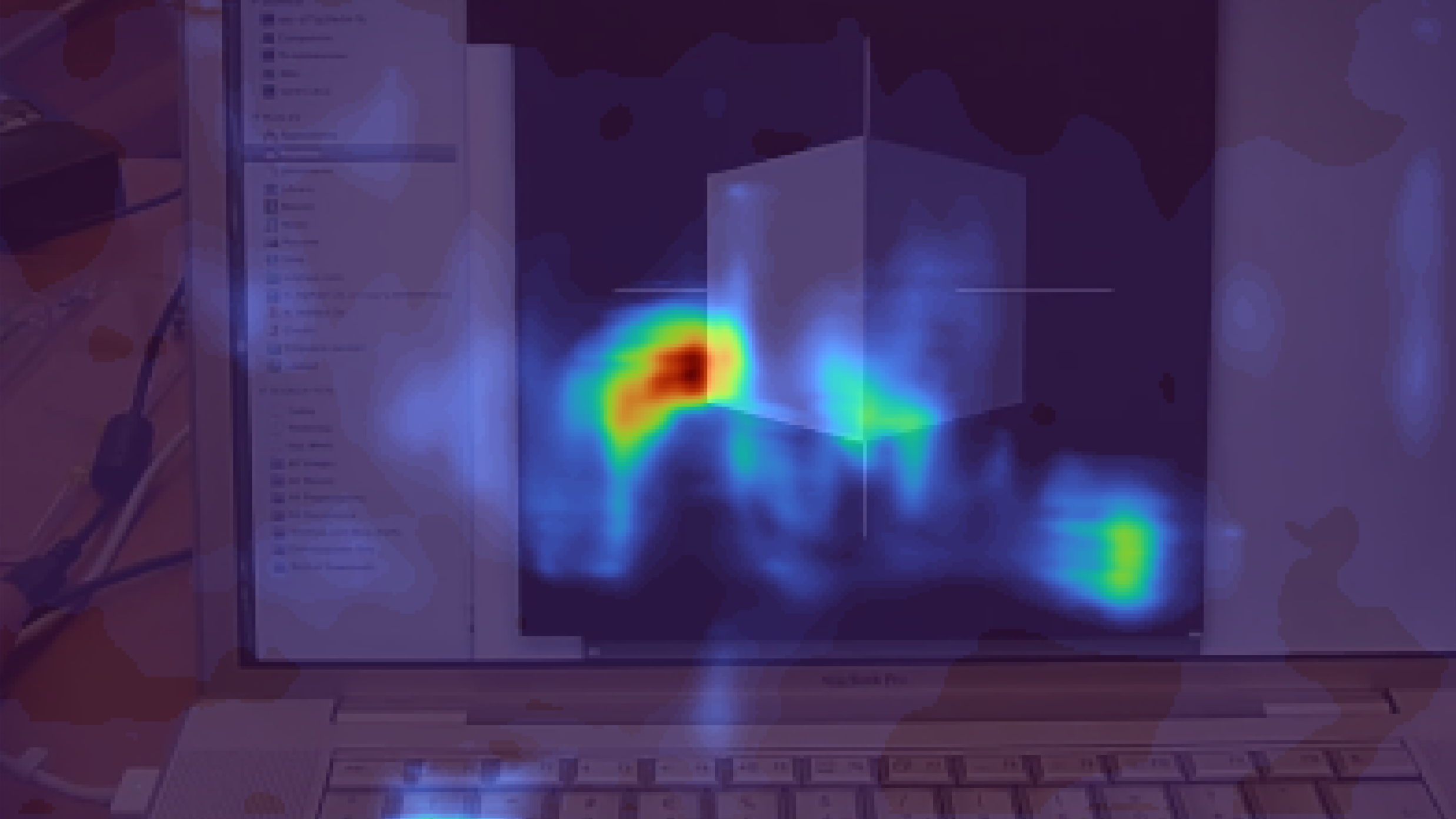} &
\tiny{Saliency shifts leftward to follow the newly revealed face as the cube turns, adapting to object motion and changing orientation.} \\ \hline
\includegraphics[width=1.5cm]{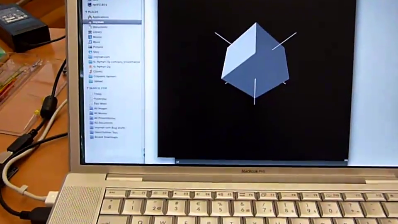} &
\includegraphics[width=1.5cm]{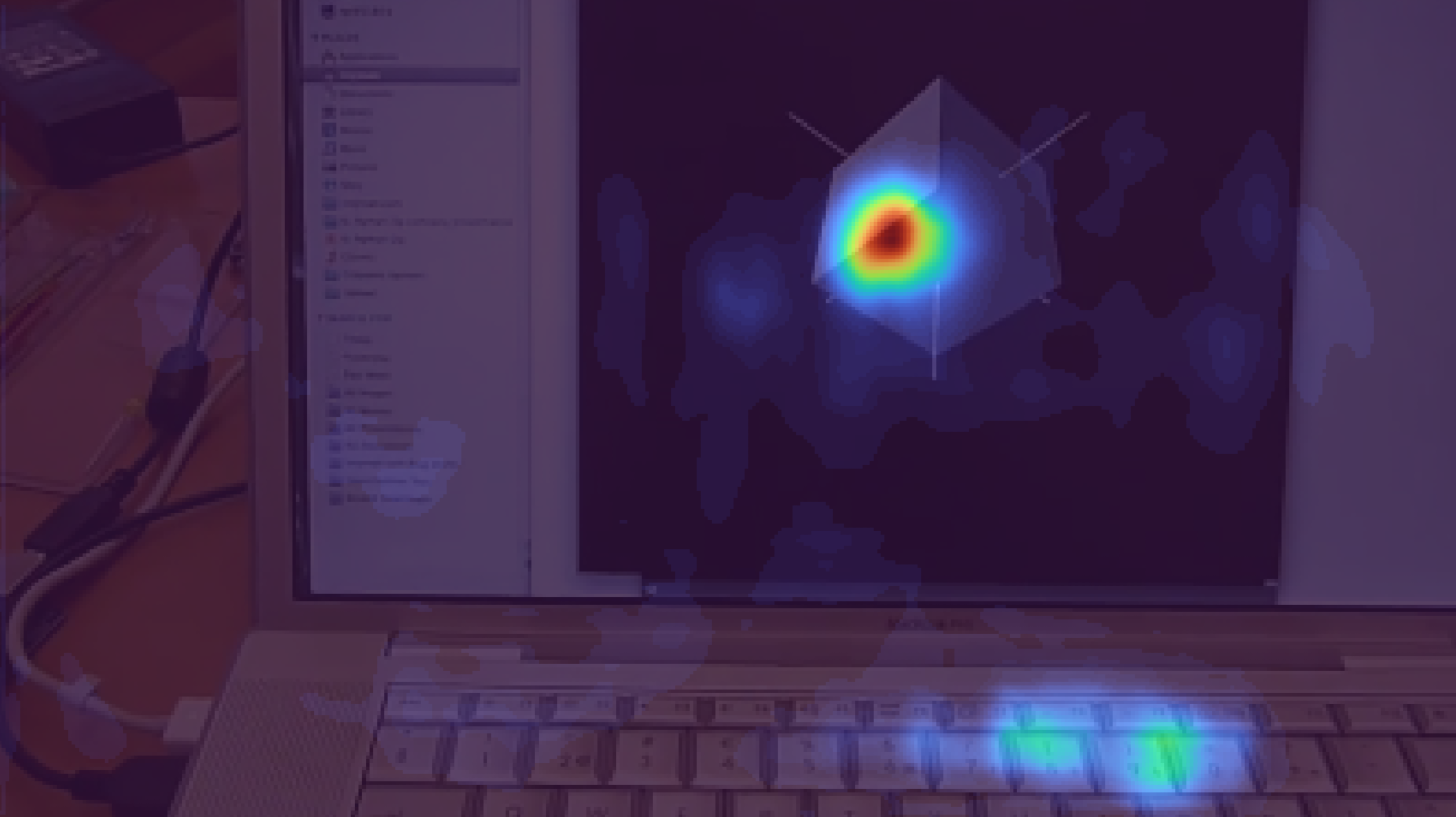} &
\tiny{Focus converges on the cube’s front as it turns to the rear, with the model adapting to the most visually prominent surface with each change.} \\ \hline
\end{tabular}
} 
\label{tab:dynsal_lsvq_konvid}
\end{table*}

\begin{figure}[tb]
    \centering \includegraphics[width=0.47\textwidth]{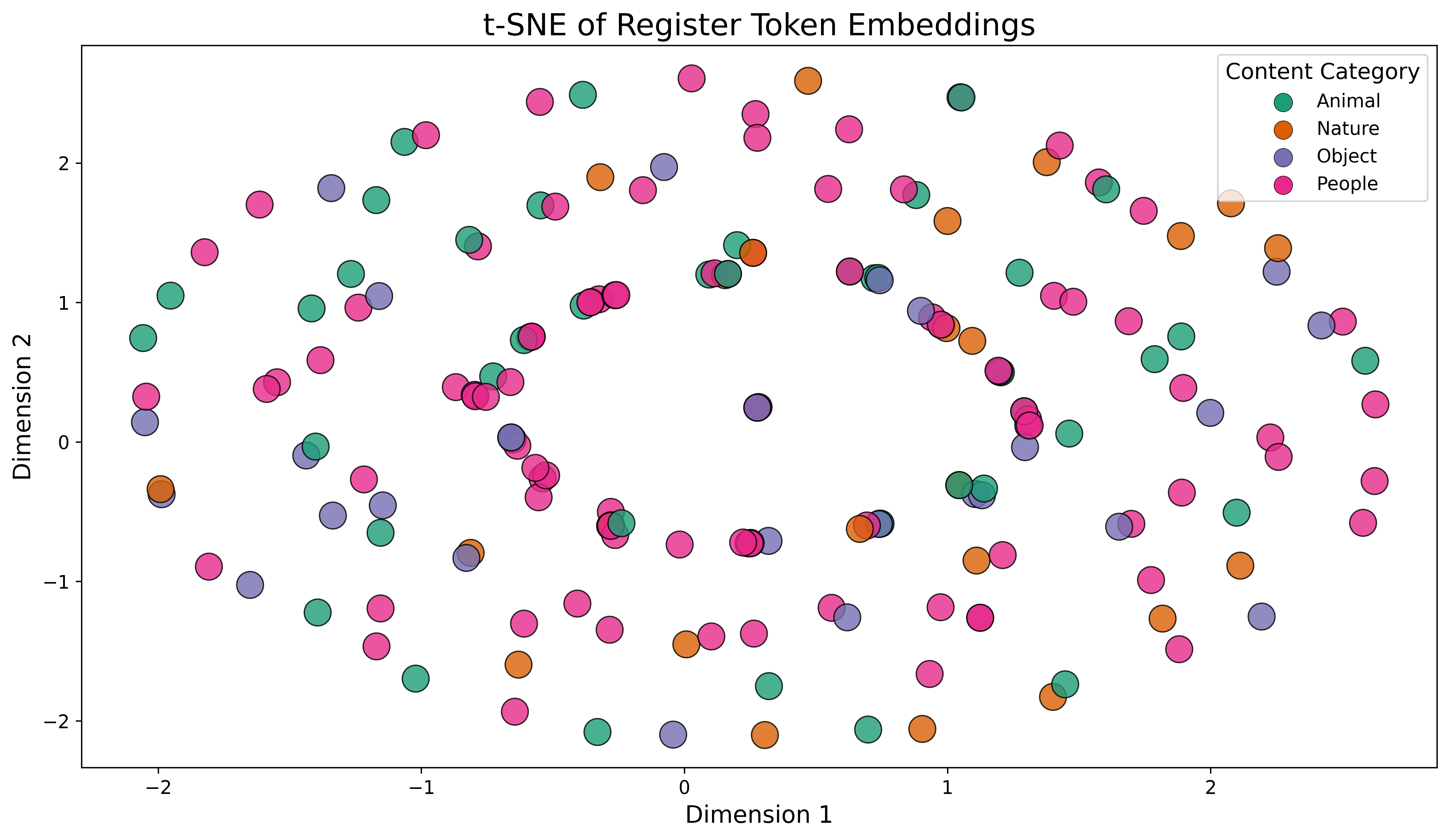}  
    \caption{t-SNE plot of Register Token Embeddings with soft grouping by content despite no semantic supervision.}
    \label{fig:reg_token_tsne}
\end{figure}

To assess the sensitivity of the global content encapsulated by the register-tokens, we extract the embeddings of all 4 register-tokens from our trained model for each video within the DHF1K dataset. To encapsulate the global context learned for each video, the mean of the token embeddings associated with that particular video is computed, resulting in a single pooled vector per video. Next, we employ t-SNE \cite{vanDerMaaten2008tsne}, a nonlinear dimensionality reduction method that transforms high-dimensional features into a two-dimensional space, ensuring the preservation of local similarities among data points. Figure \ref{fig:reg_token_tsne} illustrates a 2D t-SNE projection of the mean-pooled register-token embeddings for every video in DHF1K. Each point in the figure represents a video, with its color reflecting the type of content, as indicated in the DHF1K metadata. 
This graphic shows that videos from the same content category cluster together, even though category information was not used during training. This clustering shows that the learned register-token embeddings encode relevant content-related information, allowing for more consistent, content-aware attention in the saliency prediction module.

\begin{figure}[tb]
\centering
\includegraphics[width=1.0\linewidth]{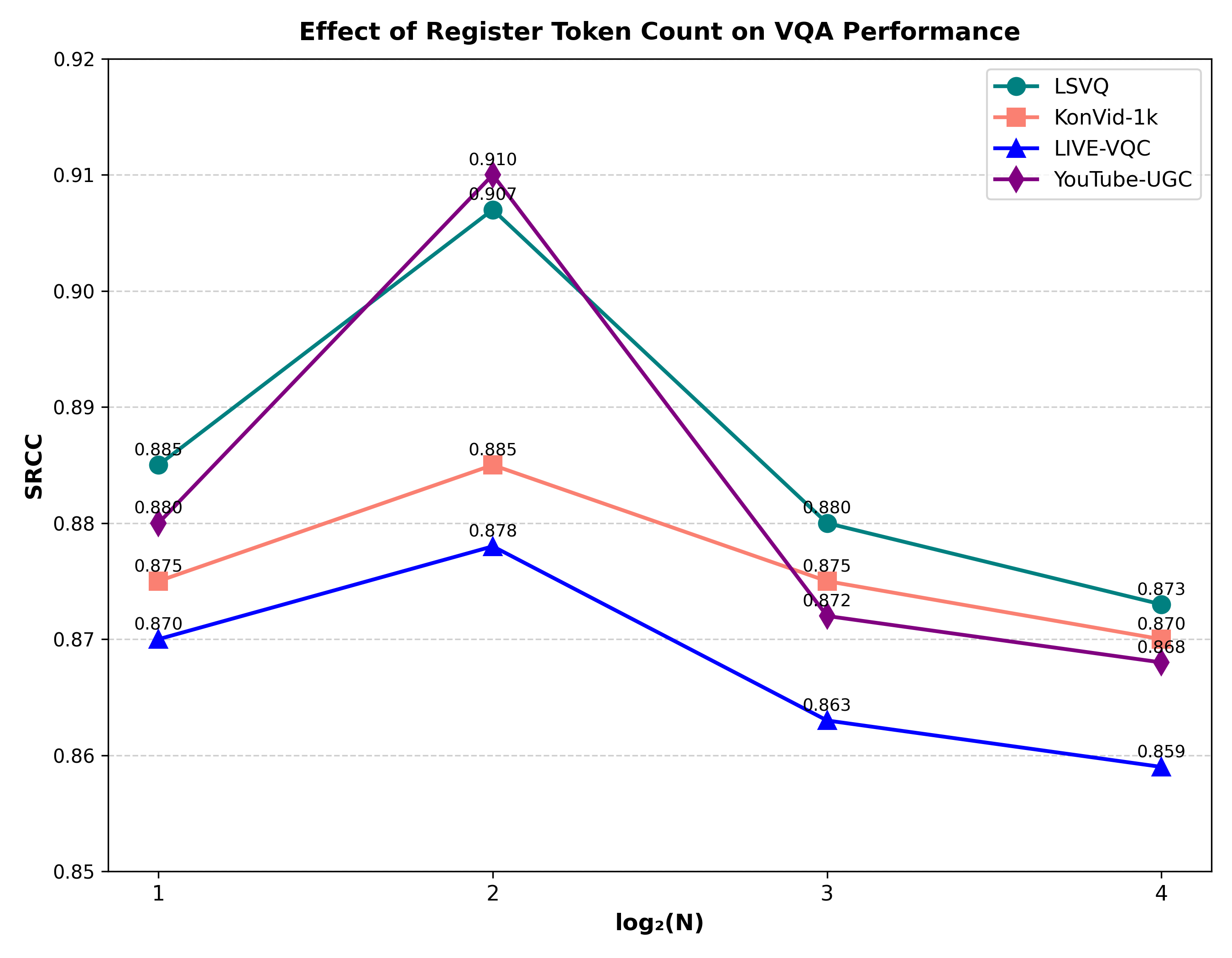}
\caption{Effect of register token count on SRCC for four datasets with $\log_2(N)$ on x-axis and SRCC on y-axis (DHF1K). SRCC peaks at $N=4$ for all cases, showing that a moderate number of register-tokens yields the best VQA performance.}
\label{fig:reg_tokens_SRCC}
\end{figure}

Next, we examine the impact of the quantity of register-tokens, denoted as $N$, on the accuracy of saliency prediction as well as the performance in downstream VQA tasks. For the VQA analysis, we vary $N \in \{2,4,8,16\}$ in the saliency backbone pretrained on DHF1K and, then, evaluate the results for the DAGR-VQA on LSVQ, KonVid-1k, LIVE-VQC, and YouTube-UGC datasets. Figure~\ref{fig:reg_tokens_SRCC} shows SRCC values plotted against $\log_2(N)$. SRCC improves from $N=2$ to $4$, but declines for $N \ge 8$, indicating moderate token numbers are advantageous, while larger sets add redundancy and reduce representation distinctiveness. We see a consistent pattern when evaluating VQA performance based on the saliency backbone itself. On DHF1K, VQA performance peaks at $N=4$ and decreases for $N \ge 8$, whereas on DIEM the best VQA performance is obtained at $N=2$, with $N=4$ remaining competitive but higher values again degrading accuracy. This supports the intuition that $N$ controls the capacity of the global prior: small $N$ encourages a compact, strongly regularized representation (favored by the smaller, less diverse DIEM), while a slightly larger $N$ is advantageous for the more diverse DHF1K dataset and for complex real-world VQA benchmarks. Overall, these results indicate that the optimal number of register-tokens lies in a low range ($N \in \{2,4\}$), which offers a good trade-off across both saliency datasets and all 4 VQA datasets.

\begin{figure}[tb]
    \centering
    \includegraphics[width=1.0\linewidth]{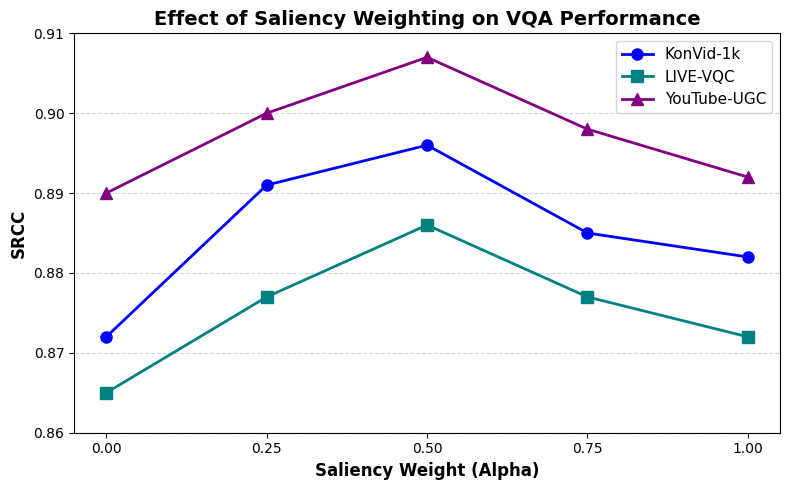}  
    \caption{Effect of saliency weighting ($\alpha$) on SRCC for three datasets, with weighting factor $\alpha$ on x-axis and SRCC on y-axis. Peak performance is achieved at $\alpha = 0.5$, highlighting that moderate saliency yields the best VQA results.}
    \label{fig:saliency_weight}
\end{figure}

We explore the impact of saliency-based weighting on quantitative VQA performance by implementing a saliency control parameter $\alpha$ in Figure \ref{fig:saliency_weight}. The range of $\alpha$ extends from $\alpha = 0$ to $\alpha = 1$, where $\alpha = 0$ indicates the use of only the original characteristics and $\alpha = 1$ means the exclusive use of the saliency-enhanced characteristics. Figure~\ref{fig:saliency_weight} shows the relationship between the saliency parameter $\alpha$ (displayed on the $\mathit{x}$ axis) and the SRCC scores (displayed on the $\mathit{y}$ axis), demonstrating the predictive accuracy of the model for different values of $\alpha$. The results indicate that increasing $\alpha$ from 0.0 (no saliency weight) to 0.5 produces a steady improvement in SRCC in all datasets, reaching a maximum for $\alpha=0.5$. However, for $\alpha > 0.5$, performance decreases consistently, implying that excessive emphasis on saliency may suppress relevant cues from less salient regions. These findings confirm that a balanced contribution, specifically at $\alpha=0.5$, helps the model take care of perceptually important regions, such as faces and moving objects, while still leveraging useful global information.

Finally, to better understand how each component (spatial analysis, temporal modeling, and saliency fusion) contributes to our final video quality model, we carried out an ablation study on the LSVQ and KonViD-1k datasets. The findings of this investigation are presented in Table~\ref{tab:spatiotemporal_ablation}, which illustrates that neither spatial nor temporal modeling independently suffices. Spatial-only and temporal-only configurations yield noticeably lower scores, especially on KonViD-1k. When we add register-token augmented dynamic saliency maps to guide spatial feature extraction, performance improves, highlighting the value of attention to important regions in each frame. However, combining all three elements (spatial, temporal, and saliency) leads to the best accuracy by a significant margin, with our full model achieving SRCC values of 0.892 (LSVQ) and 0.863 (KonViD-1k), and top PLCC scores on both. This progression demonstrates that each module addresses a unique aspect of video quality, and integrating all 3 is essential for the most reliable assessment results.

\paragraph{Limitations and Failure Cases.}
Our approach performs well in scenes with clear subject motion or moderate interaction. However, there are certain performance limitations. In mostly static scenarios or those with minor spatial shift, projected saliency maps tend to remain concentrated and show little dynamic variation from frame to frame, providing little advantage over simpler static approaches. In contrast, in scenes with several moving subjects or complex activity, the model may occasionally create fractured or dispersed saliency regions, highlighting several places at the same time or presenting minor frame-to-frame inconsistencies. Our ablation study also indicates that the quality of the saliency maps and the overall performance of the model is responsive to the number of registration tokens $N$ and the saliency control parameter $\alpha$. Careful selection or adaptation of these parameters might be beneficial in minimizing these limitations.


\begin{table}[ht]
\centering
\caption{Ablation study on spatio-temporal VQA analyzer components evaluated on LSVQ and KonViD-1k validation sets. Each configuration isolates or combines key modules to demonstrate their individual and collective contributions to video quality prediction.}
\setlength{\tabcolsep}{2pt}
\begin{tabular}{lccccc}
\toprule
\textbf{Configuration} & \multicolumn{2}{c}{\textbf{LSVQ}} & \multicolumn{2}{c}{\textbf{KonViD-1k}} \\
\cmidrule(lr){2-3} \cmidrule(lr){4-5}
& \textbf{SRCC} & \textbf{PLCC}  & \textbf{SRCC} & \textbf{PLCC}  \\
\midrule
Spatial Only  & 0.825 & 0.832 & 0.813 & 0.826 \\
Temporal Only & 0.872 & 0.877 & 0.838 & 0.870 \\
Spatial + Saliency Fusion & 0.865 & 0.871 & 0.820 & 0.866 \\
\textbf{Spatial + Temporal + Saliency} & \textbf{0.892} & \textbf{0.907} & \textbf{0.863} & \textbf{0.896} \\
\bottomrule
\end{tabular}
\label{tab:spatiotemporal_ablation}
\end{table}

\section{Conclusion}
We present DAGR-VQA, a novel no-reference video quality assessment framework that incorporates register-tokens as global priors alongside a convolutional saliency backbone for effective dynamic saliency prediction. Our results indicate that this strategy can improve temporally adaptive saliency modeling and enhance the ability of the model to identify perceptual degradations across various types of video content. Experimental tests on multiple public video quality datasets show that DAGR-VQA performs competitively in accuracy and has good generalizability. Ablation studies reveal that an empirical tuning of register-tokens might lead the model to focus more reliably on the most relevant regions and moments in the video sequence. The efficiency of DAGR-VQA and its modular structure facilitate rapid integration for real-time video quality monitoring in streaming system frameworks. Future work might explore adaptive saliency fusion methods and broader real-world applications.


\bibliographystyle{ACM-Reference-Format}
\bibliography{refs}

\end{document}